\documentclass[sigconf]{acmart}
\usepackage{wrapfig}
\usepackage{fontawesome5}
\usepackage{multicol}
\usepackage{multirow}
\AtBeginDocument{%
  }

\setcopyright{acmlicensed}
\copyrightyear{2026}
\acmYear{2026}
\acmConference[ICCAD '26]{IEEE/ACM International Conference on Computer-Aided Design}{November 8--12, 2026}{San Jose, CA, USA}

\begin{document}

\title{DiffAxE: Diffusion-driven Hardware Accelerator
Generation and Design Space Exploration}

\author{Arkapravo Ghosh}
\email{arkaprav@usc.edu}
\affiliation{%
  \institution{University of Southern California}
  \city{Los Angeles}
  \state{California}
  \country{USA}
}

\author{Abhishek Moitra}
\email{abhishek.moiitra@yale.edu}
\affiliation{%
  \institution{Yale University}
  \city{New Haven}
  \state{Connecticut}
  \country{USA}
}

\author{Abhiroop Bhattacharjee}
\email{abhiroop.bhattacharjee@yale.edu}
\affiliation{%
  \institution{Yale University}
  \city{New Haven}
  \state{Connecticut}
  \country{USA}
}

\author{Ruokai Yin}
\email{ruokai.yin@yale.edu}
\affiliation{%
  \institution{Yale University}
  \city{New Haven}
  \state{Connecticut}
  \country{USA}
}

\author{Priyadarshini Panda}
\email{priya.panda@usc.edu}
\affiliation{%
  \institution{University of Southern California}
  \city{Los Angeles}
  \state{California}
  \country{USA}
}
%

\begin{abstract}
  Design space exploration (DSE) for hardware accelerators faces intractable search spaces of $O(10^{17})$ configurations with non-convex, many-to-one hardware--performance mappings. Conventional search methods such as Bayesian optimization, reinforcement learning and genetic algorithms suffer from slow iterative sampling, while deep learning classification-based approaches are limited to $O(10^3)$ spaces. We propose DiffAxE, a diffusion based generative DSE framework that models accelerator design as 1-D image synthesis conditioned on target performance, efficiently capturing non-differentiable, non-bijective hardware--performance relationships. Our method achieves 0.86\% lower error than Bayesian optimization with a $17{,}000\times$ search time speedup for the optimal hardware configuration, and 30\% lower error than the generative DSE work, GANDSE at $1.83\times$ search time overhead. 
  Applied to LLM inference, the hardware configurations generated from DiffAxE achieve $3.37\times$ and $7.75\times$ EDP reduction on a 32nm ASIC and Xilinx Ultrascale+ VU13P FPGA, respectively, over the state-of-the-art DOSA framework.

\end{abstract}


\begin{teaserfigure}
  \includegraphics[width=\textwidth]{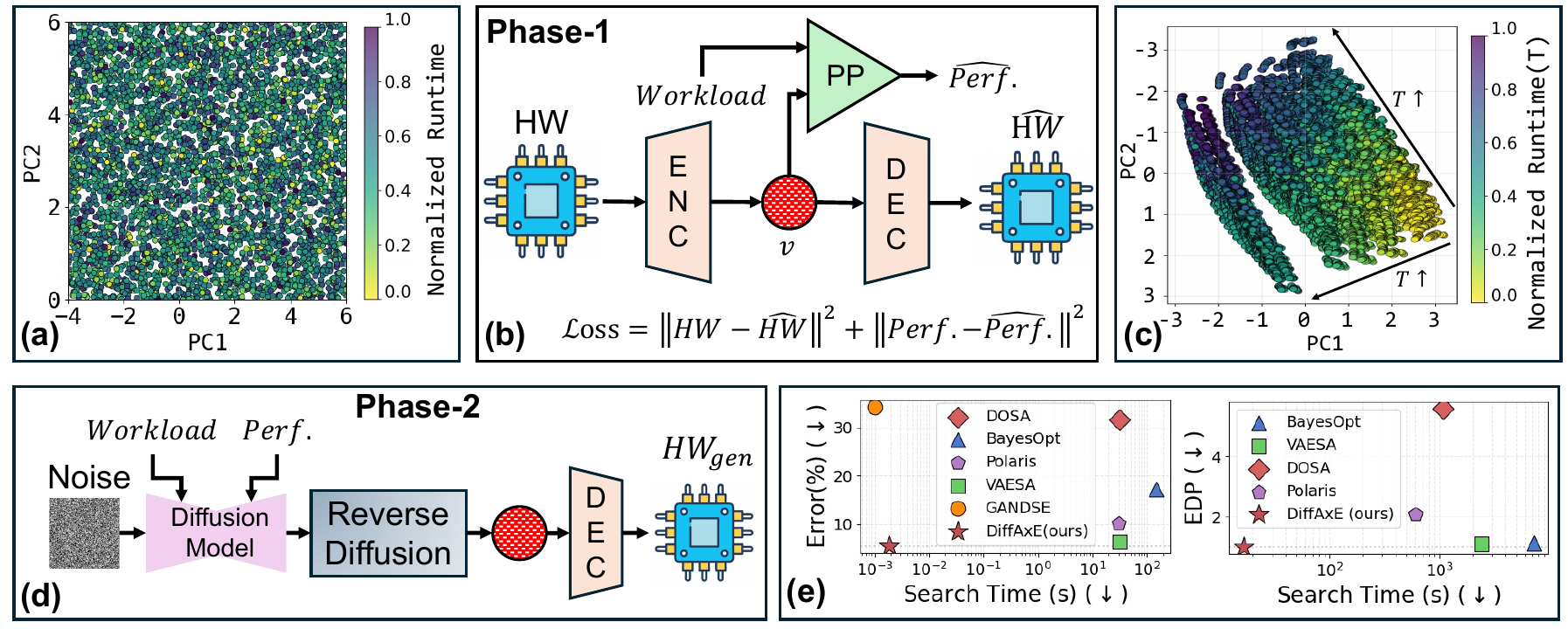}
  \caption{DiffAxE overview. (a) Principal component analysis (PCA) of the raw hardware-performance landscape reveals a highly irregular and discontinuous runtime distribution. (b) Phase 1: Joint training of an autoencoder (AE) and performance predictor (PP) structures the latent space by performance, enabling a generative model to learn the hardware-performance mapping. (c) Phase 1 latent space $(v)$ PCA: a smooth, structured space where runtime varies continuously with latent hardware configurations. (d) Phase 2: A conditional diffusion model trained on the smooth latent space generates target performance and workload-conditioned hardware via reverse diffusion. (e) DiffAxE achieves the lowest generation error and energy-delay-product (EDP) at orders of magnitude smaller search time, outperforming prior works: BayesOpt \cite{reagen2017case}, DOSA \cite{hong2023dosa}, VAESA \cite{huang2022learning}, Polaris \cite{sakhuja2024polaris} and GANDSE \cite{feng2023gandse}.}
  \label{fig:teaser}
\end{teaserfigure}

\newcommand{\ourwork}{DiffAxE}
\maketitle

\renewcommand{\thefootnote}{}%
\footnotetext{This work was supported in part by CoCoSys, a JUMP2.0 center sponsored by DARPA and SRC, the National Science Foundation (CAREER Award, Grant \#2312366, Grant \#2318152), the DARPA Young Faculty Award, the DoE MMICC center SEA-CROGS (Award \#DE-SC0023198) and the Global Industrial Technology Cooperation Center(GITCC) program.}%
\renewcommand{\thefootnote}{\arabic{footnote}}%

\section{Introduction}
The growing complexity of AI workloads across cloud and edge platforms has driven demand for application-specific hardware optimized for power-performance-area (PPA) tradeoffs in deep-neural network (DNN) inference \cite{zhao2023neural, sze2017efficient}. This necessitates efficient design space exploration (DSE) tools that rapidly generate hardware architectures satisfying tight PPA constraints. Typical DNN inference accelerators are specialized processors \cite{zhao2023neural, reagen2021weightless}, typically comprising a systolic array of multiply-and-accumulate (MAC) units with on-chip SRAM buffers performing tiled matrix multiplications while streaming data to and from off-chip DRAM.
\begin{figure}[h]
    \centering
    \includegraphics[width=0.48\textwidth]{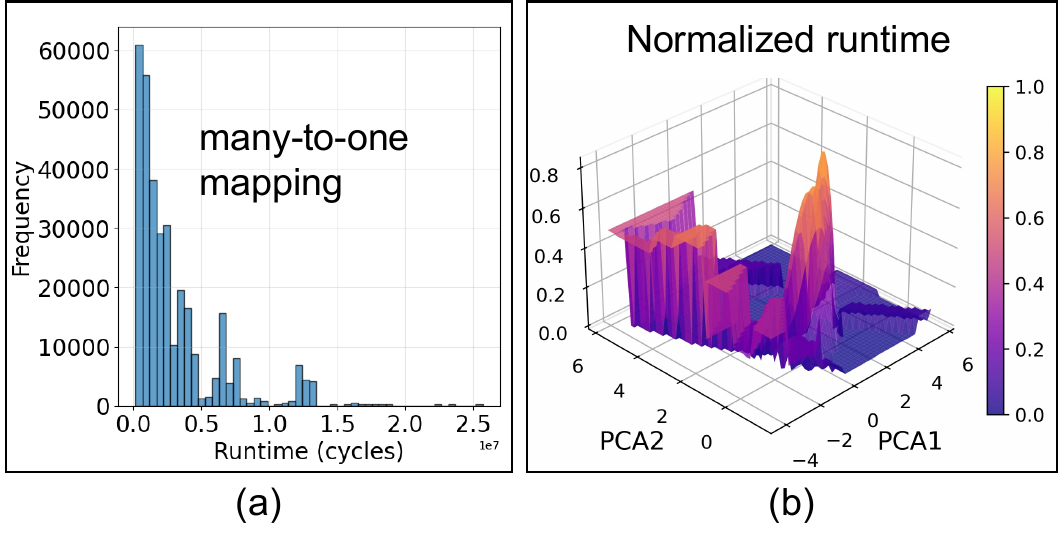}
    \caption{(a) Many-to-one mapping of hardware configuration to runtime: different hardware designs achieve the same runtime for executing the QKV layer of DeiT-B model. (b) PCA of the raw hardware-performance landscape reveals a highly irregular and discontinuous runtime distribution.}
    \label{fig:dse_challenge}
\end{figure}
As illustrated in Fig.~\ref{fig:teaser}(a), the hardware-performance (i.e. runtime) mapping exhibits no discernible structure or relation in the raw space -- revealing two fundamental challenges that make DSE intractable. First, the mapping is \textbf{many-to-one} as shown in Fig.\ref{fig:dse_challenge}(a): multiple distinct hardware configurations yield identical runtime, precluding direct inverse derivation of designs achieving a target performance. Second, as shown in Fig.~\ref{fig:dse_challenge}(b), the mapping is \textbf{non-differentiable and discontinuous}, ruling out gradient-based \cite{hong2023dosa, liu2018darts, wu2019fbnet} and differential approximation \cite{hong2023dosa} methods. Together, these properties make DSE fundamentally a \textit{conditional generation} problem: given a target performance and workload, one must directly generate hardware configurations satisfying the target, rather than iteratively searching the design space or relying on approximate differentiable models of the non-differentiable hardware-performance function. 

Iterative black-box methods — Bayesian optimization (BO) \cite{reagen2017case, huang2022learning}, gradient descent (GD) \cite{hong2023dosa, sakhuja2024polaris}, reinforcement learning (RL) \cite{kao2020confuciux}, and genetic algorithms (GA) \cite{kao2020gamma, kao2022digamma} — require 3--10 hours of repeated simulator calls, while deep learning (DL)-based recommendation models \cite{samajdar2021airchitect, seo2025airchitect} are confined to impractically small design spaces. The only generative prior work, GANDSE \cite{feng2023gandse}, enables direct hardware generation within milliseconds, but approximates the non-differentiable performance function with a differentiable surrogate — an assumption fundamentally violated by the irregular hardware-performance landscape (Fig.~\ref{fig:dse_challenge}(b)) — yielding 35\% average generation error for optimal hardware configurations. \textit{Hence, existing methods face a fundamental tradeoff: iterative search methods are prohibitively slow, while faster generative approaches suffer from high generation error due to invalid assumptions about the performance landscape.}

Since the hardware-to-performance mapping is many-to-one, recovering hardware configurations that achieve a target performance for a given workload has no unique solution, yet this is precisely what efficient DSE demands. Diffusion models are naturally suited to this: rather than learning a single inverse, they learn a \textit{distribution} over all hardware configurations consistent with a target, inherently handling the non-invertibility. Moreover, they have demonstrated remarkable success in conditional generation \cite{ho2020denoising, rombach2022high, dhariwal2021diffusion}, producing high-quality diverse outputs conditioned on arbitrary signals — without requiring the conditioning signal to be differentiable — at millisecond-scale inference times. \textit{Hence, we envisage that diffusion models can be used for rapid DSE, solving the long search time issue with iterative methods \cite{huang2022learning, reagen2017case, sakhuja2024polaris, hong2023dosa}.}

However, as shown in Fig.~\ref{fig:teaser}(a), principal component analysis (PCA) of the runtime reveals no performance correlated structure—points with similar runtime are scattered arbitrarily. Unlike in image generation where samples of the same class share natural visual similarity, hardware configurations achieving similar performance are arbitrarily scattered in the raw design space (Fig.~\ref{fig:teaser}(a)), providing no exploitable structure for a generative model. 
Thus, latent diffusion \cite{rombach2022high} that can encode hardware configurations into a \textit{performance-aware} latent space is a strict necessity here to render reliable conditional generation. 

Based on the above insights, we propose \textbf{\ourwork{}}: the first diffusion model-based generative DSE framework for AI accelerators. \ourwork{} comprises of two phases. As shown in Fig.\ref{fig:teaser}(b), Phase-1 learns a performance-aware latent space (Fig. \ref{fig:teaser}(c)) that imposes structure on the latent hardware-performance landscape, enabling accurate performance-conditioned generation. Phase-2 (Fig.~\ref{fig:teaser}(d)) trains a conditional diffusion model on this latent space, inheriting the millisecond-scale inference speed of diffusion models. Together, as shown in Fig.~\ref{fig:teaser}(e), \ourwork{} achieves the lowest generation error and EDP at millisecond-scale search time, outperforming all baselines. Our contributions are as follows:

\begin{enumerate}
    \item 
    \ourwork{} generates hardware designs within $\pm 5.45\%$ of the target objective while exploring design spaces on the order of $10^{17}$. 
    We evaluate our approach on 600 deep neural network (DNN) workloads, significantly exceeding prior work, which typically considers around 70 workloads~\cite{huang2022learning, sakhuja2024polaris, hong2023dosa}. 

    \item \ourwork{} achieves on an average 9.8\% lower energy-delay product (EDP) than prior  DSE tools~\cite{huang2022learning}, while performing the search $145.6\times$ faster. Specifically, for large language model (LLM) inference workloads, it achieves up to $2.07\times$ and $5.6\times$ lower EDP compared to Polaris~\cite{sakhuja2024polaris} and DOSA~\cite{hong2023dosa}, respectively.

    \item \ourwork{} achieves up to 93\% lower runtime compared to generative model based DSE tool GANDSE\cite{feng2023gandse}. In addition, our generative approach uses 32\% fewer parameters than AIRCHITECT-v2~\cite{seo2025airchitect}- a regression based tool. 
\end{enumerate}

\section{Related Work}
\label{sec:related}

\noindent\textbf{Black-box Optimization Methods:}
Reagen et al. \cite{reagen2017case} first demonstrated Bayesian optimization (BO) for accelerator DSE, iteratively building a statistical surrogate of the performance function to guide search. VAESA \cite{huang2022learning} and Polaris \cite{sakhuja2024polaris} improve upon this by learning a variational autoencoder (VAE)-based latent space and applying BO and gradient descent (GD) within it respecitively. DOSA \cite{hong2023dosa} constructs a differentiable analytical performance model to enable gradient-based one-loop co-search of hardware and mapping. ConfuciuX \cite{kao2020confuciux} employs RL to search the compute and buffer sizes for a fixed dataflow, while GAMMA \cite{kao2020gamma} applies genetic algorithms to optimize mapping for a fixed hardware. Despite their differences, all these methods rely on iterative simulator evaluations, incurring search times of up to several hours, and gradient-based approaches additionally assume differentiability of the performance function — an assumption violated by the highly discontinuous hardware-performance landscape (Fig.~\ref{fig:dse_challenge}(b)).

\noindent\textbf{Deep-Learning-based Methods:}
AIRCHITECT \cite{samajdar2021airchitect} formulates design selection as multi-class classification, while AIRCHITECT-v2 \cite{seo2025airchitect} addresses scalability through latent embeddings and hybrid classification-regression. Nevertheless, both are limited to design spaces of only 768 configurations encompassing two hardware parameters, and cannot generate performance-conditioned configurations.

\noindent\textbf{Generative Methods:}
GANDSE \cite{feng2023gandse} is the only prior work to directly generate hardware configurations, using a GAN-based approach with a differentiable surrogate of the performance function. While it achieves millisecond generation times, approximating the non-differentiable performance function with a surrogate yields 35\% average error (see Sec. 6.1, Table \ref{tab:baseline_exp1}), and is based on a smaller ($O(10^6)$) design space.

\section{Background}
\label{sec:bgnd}

\subsection{DSE Problem Formulation}
We model AI workloads as general matrix multiply (GEMM) operations: $(M,K)\times(K,N)=(M,N)$, which dominate computation in large-language models (LLMs) and Vision Transformers (ViTs), the state-of-the-art DNNs. Performance $p = f(HW, w)$, which may refer to any hardware efficiency metric such as inference runtime, power, energy, or energy-delay product (EDP), depends jointly on the hardware configuration parameters $HW$ and workload $w$. $HW$ is parameterized by seven key variables—array size, loop order, buffer sizes, and DRAM bandwidth (Table~\ref{tab:design_parameters}). The resulting design space $|\mathcal{D}| = |R|\times|C|\times|IPSz|\times|WTSz|\times|OPSz|\times|BW|\times|Loop Order|$ spans $O(10^{13\text{-}17})$ configurations \cite{huang2022learning}. In this work, a workload $w$ is defined by the GEMM dimensions, i.e., $w = (M, K, N) \in \mathbb{R}^3$.

\begin{table}[h]
\caption{Typical accelerator design parameters}
\centering
\resizebox{\columnwidth}{!}{%
\begin{tabular}{ll}
\hline
\multicolumn{1}{c}{\textbf{Design Parameter}} & \multicolumn{1}{c}{\textbf{Typical Range}} \\
\hline
MAC array size (R, C) & 4 -- 128 (integers only) \\
Buffer Size (IPSz, WTSz, OPSz) & 4--1024 kB (step: 128 B) \\
DRAM Bandwidth ($BW$) & 2 -- 32 B/cycle (step: 1 B/cycle) \\
Loop Order & mnk, nmk, knm, nkm, mkn, kmn\\
\textbf{Total design space} & $1.5 \times 10^{18}$\\
\hline
\textbf{Workload, $\mathbf{w}$} & (M, K, N) $w\in\mathbb{R}^3$ \\\hline
\end{tabular}%
}
\label{tab:design_parameters}
\end{table}

\subsection{Diffusion Models}
Diffusion models learn data distributions via a forward noising process and a learned reverse denoising process.

\noindent \underline{\textbf{Forward diffusion:}}
A data sample $\mathbf{x}_0 \sim p_{\text{data}}(\mathbf{x})$ is progressively perturbed over $T$ steps with Gaussian noise:
\begin{equation}
\mathbf{x}_t = \sqrt{\bar{\alpha}_t}\,\mathbf{x}_0 + \sqrt{1-\bar{\alpha}_t}\,\boldsymbol{\epsilon}, \quad \boldsymbol{\epsilon}\sim\mathcal{N}(\mathbf{0},\mathbf{I}),
\end{equation}
where $\bar{\alpha}_t=\prod_{s=1}^t \alpha_s$, $\alpha_t=1-\beta_t$, and $\{\beta_t\}$ is a predefined noise schedule.

\noindent \underline{\textbf{Training objective:}}
The model is trained to predict the injected noise using a denoising objective:
\begin{equation}
\mathcal{L}_{DDM} = \left\|\boldsymbol{\epsilon}-\boldsymbol{\epsilon}_\theta(\mathbf{x}_t,t)\right\|^2,
\end{equation}
where $\boldsymbol{\epsilon}_\theta$ denotes the predicted noise.

\noindent \underline{\textbf{Reverse diffusion:}}
The reverse process reconstructs data by iteratively denoising:
\begin{equation}
p_\theta(\mathbf{x}_{t-1}|\mathbf{x}_t)=\mathcal{N}(\boldsymbol{\mu}_\theta(\mathbf{x}_t,t),\sigma_t^2\mathbf{I}),
\end{equation}
with mean
\begin{equation}
\boldsymbol{\mu}_\theta(\mathbf{x}_t,t)=\frac{1}{\sqrt{\alpha_t}}
\left(\mathbf{x}_t-\frac{1-\alpha_t}{\sqrt{1-\bar{\alpha}_t}}\boldsymbol{\epsilon}_\theta(\mathbf{x}_t,t)\right).
\end{equation}
Sampling starts from $\mathbf{x}_T\sim\mathcal{N}(\mathbf{0},\mathbf{I})$ and iteratively denoises to obtain $\mathbf{x}_0$:
\begin{equation}
    \mathbf{x}_{t-1} = \mu_\theta(x_t, t) + \sigma_t \mathbf{z}, \quad \mathbf{z} \sim 
    \begin{cases}
        \mathcal{N}(\mathbf{0}, \mathbf{I}) & \text{if } t > 1, \\
        \mathbf{0} & \text{if } t = 1.
    \end{cases}
\end{equation}
Motivated by the success of diffusion models in conditional image generation \cite{ho2020denoising, rombach2022high, dhariwal2021diffusion}, we encode accelerator configurations as 1-D sequences and employ a performance-conditioned denoising diffusion model (DDM) to generate hardware designs, enabling efficient generative design space exploration.

\section{Methodology}

\begin{figure*}[h]
    \centering
    \includegraphics[width=\textwidth]{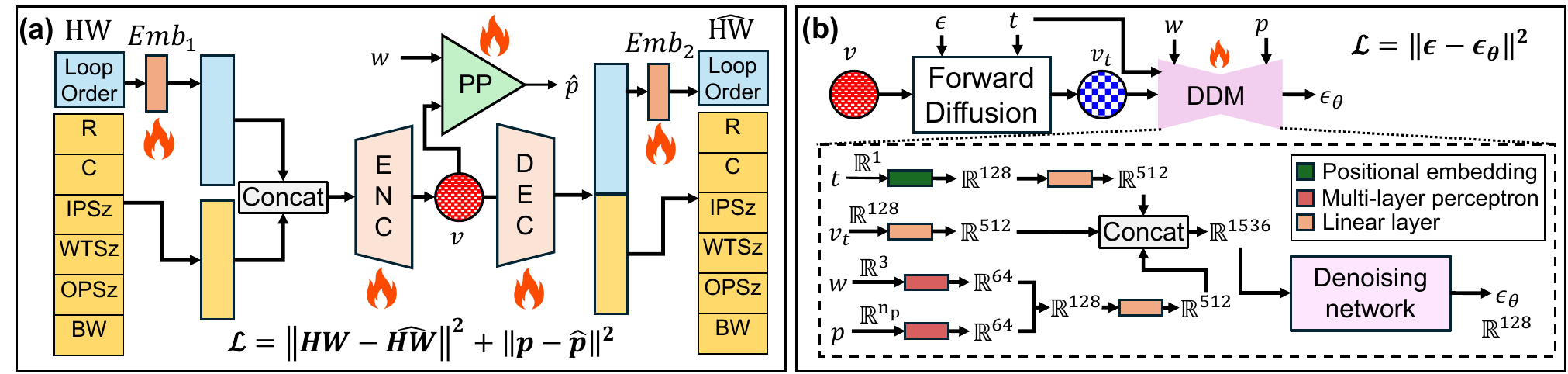}
    \caption{(a) Phase-1: performance-guided encoding of hardware designs into a latent space. The AE ensures reconstruction to the original space while the PP organizes the latent space to encourage lower Euclidean distance between latent vectors with similar performances. \faFire{} denotes the trainable modules. (b) Phase-2: training the DDM conditioned on the workload $w$ and the performance $p$ to predict the noise injected in the noisy latent $v_{t}$ during forward diffusion. The inset shows how the timestep ($t$), $v_{t}$ and the conditioning signals are processed before passing to the denoising network.}
    \label{fig:method}
\end{figure*}

\ourwork{} reformulates DSE as a conditional generation problem: rather than inverting the non-bijective mapping $f$, we learn the conditional distribution $f(HW \mid p, w)$ using a diffusion model. However, training directly on the raw hardware space is ineffective — hardware configurations with similar performance are arbitrarily scattered (Fig.~\ref{fig:teaser}(a)), providing no exploitable structure for the model. Motivated by latent diffusion \cite{rombach2022high} and prior latent-space DSE works \cite{huang2022learning, sakhuja2024polaris}, we first learn a performance-aware latent space where similar-performance designs cluster together, then train a conditional DDM on this structured space. Training proceeds in two phases (Fig.~\ref{fig:method}): Phase~1 ($\phi_1$) constructs the latent space via a jointly trained autoencoder (AE) and performance predictor (PP); Phase~2 ($\phi_2$) trains the DDM conditioned on target performance $p$ and workload $w$. At inference, the DDM generates latent vectors satisfying target performance constraints, decoded into valid hardware configurations via the AE decoder.

\subsection{Phase 1: Performance-guided Latent Encoding}

The goal of Phase~1 is to construct a latent space that is simultaneously \textit{reconstructible} — so that any latent point can be mapped back to a valid hardware configuration — and \textit{performance-aware} — so that hardware designs with similar performance are geometrically proximate. This structure is essential: without it, the diffusion model has no meaningful signal to associate regions of the latent space with specific performance targets. As shown in Fig.~\ref{fig:method}(a), each accelerator design is represented as a 7-dimensional vector (Table~\ref{tab:design_parameters}). The categorical loop order is embedded into an 8-dimensional continuous vector via a learnable embedding ($\text{Emb}_1$) and concatenated with the remaining numerical features to form a 14-dimensional input. A 3-layer MLP encoder (ENC) maps this to a 128-dimensional latent vector $v$:
$\text{Linear}(14,512)\!\to\!\text{Linear}(512,256)\!\to\!\text{Linear}(256,128),$
with a symmetric decoder (DEC) reconstructing the original design, trained with reconstruction loss:
\begin{equation}
    \mathcal{L}_{\text{recon}} = \|HW - \hat{HW}\|^2.
\end{equation}
Reconstruction loss alone, however, imposes no semantic structure on the latent space — nearby latent points may correspond to hardware designs with vastly different performance. To address this, a performance predictor (PP) is trained jointly with the AE, encouraging the encoder to organize the latent space by performance. The PP has two branches that capture the two sources of performance variation: (1) a workload processor — a 4-layer MLP mapping $w \in \mathbb{R}^3$ to a scalar, capturing how workload dimensions affect runtime; and (2) a latent performance processor — a linear layer projecting $v$ to a scalar, capturing how the hardware configuration affects runtime. The predicted performance $\hat{p}$ is their sum, trained via MSE loss:
\begin{equation}
    \mathcal{L}_{\text{pred}} = \|p - \hat{p}\|^2.
\end{equation}
Here, $p, \hat{p} \in \mathbb{R}^{n_p}$, with $n_p = 1$ denoting a single-objective formulation, while $n_p > 1$ generalizes to multi-objective targets (e.g., joint runtime and power), as further detailed in Section~\ref{sec:edp_opt}. The AE and PP are jointly optimized with a combined loss:
\begin{equation}
    \mathcal{L}_{\text{total}} = \mathcal{L}_{\text{recon}} + \mathcal{L}_{\text{pred}},
\end{equation}
where $\mathcal{L}_{\text{recon}}$ ensures the latent space remains reconstructible while $\mathcal{L}_{\text{pred}}$ imposes performance-correlated geometric structure. The result, shown in Fig.~\ref{fig:teaser}(c), is a smooth latent space where runtime varies continuously along interpretable directions — a stark contrast to the irregular raw landscape of Fig.~\ref{fig:dse_challenge}(b). 

\subsection{Phase 2: Conditional Diffusion Model Training}

With a structured latent space $v$ in place, Phase~2 trains a conditional DDM to learn the distribution of latent hardware vectors $v$ given target performance $p$ and workload $w$. Operating in the latent space rather than the raw hardware space is critical: the smooth performance-correlated structure from Phase~1 enables the diffusion model to reliably navigate toward target-conditioned regions during reverse diffusion. As shown in Fig.~\ref{fig:method}(b), the DDM is trained to predict the noise $\boldsymbol{\epsilon}$ injected into noisy latent vectors $v_t$, conditioned on timestep $t$, workload $w$, and performance $p$. It comprises of two sub-modules:

\noindent\textbf{Signal Processor:} The timestep, performance $p$, and workload $w$ are processed in parallel and fused before denoising. Sinusoidal time embeddings (dim 128), $p$ and $w$ processed by separate two-layer MLPs and concatenated, and noisy latent $v_t \in \mathbb{R}^{128}$ are each projected to dim 512 and concatenated to form a 1536-dim unified representation.

\noindent\textbf{Denoising Network:} An asymmetric U-Net processes the 1536-dim input through a downsampling MLP path ($1536\!\to\!256$, LayerNorm, ReLU, dropout), a 256-dim bottleneck, and an upsampling path with skip connections back to dim 512, generating $\boldsymbol{\epsilon}_\theta \in \mathbb{R}^{128}$. Skip connections preserve fine-grained latent structure across the denoising path. Training follows \cite{ho2020denoising} with a linear $\beta$-schedule over $T=1000$ steps, minimizing MSE between predicted $\epsilon_{\theta}$ and true noise $\epsilon$.

\subsection{Target-conditioned Hardware Generation}

At inference, as illustrated in Fig. \ref{fig:teaser}(d), the trained DDM generates hardware configurations conditioned on a target performance $p$ and workload $w$ via reverse diffusion. Starting from $v_T \sim \mathcal{N}(0,\mathbf{I})$, the DDM iteratively denoises over $T (=1000)$ steps via Eqs.~(4)--(5), navigating through the performance-structured latent space toward the target-conditioned region. The performance-aware latent space from Phase~1 is essential here: it ensures that the reverse diffusion trajectory converges to hardware configurations genuinely satisfying the target performance, rather than arbitrary points in the latent space. At $t{=}1$, the frozen AE decoder (Fig.~\ref{fig:teaser}(d)) maps the denoised latent back to the design space, where outputs are inverse-normalized and rounded to the nearest valid discrete states.

\subsection{DSE for EDP Optimization}
\label{sec:edp_opt}

So far, we have considered conditioning on a single performance metric (runtime). We now extend \ourwork{} to energy-delay product (EDP) optimization by jointly conditioning
\begin{wrapfigure}{r}{0.6\linewidth}
    \centering
    \includegraphics[width=\linewidth]{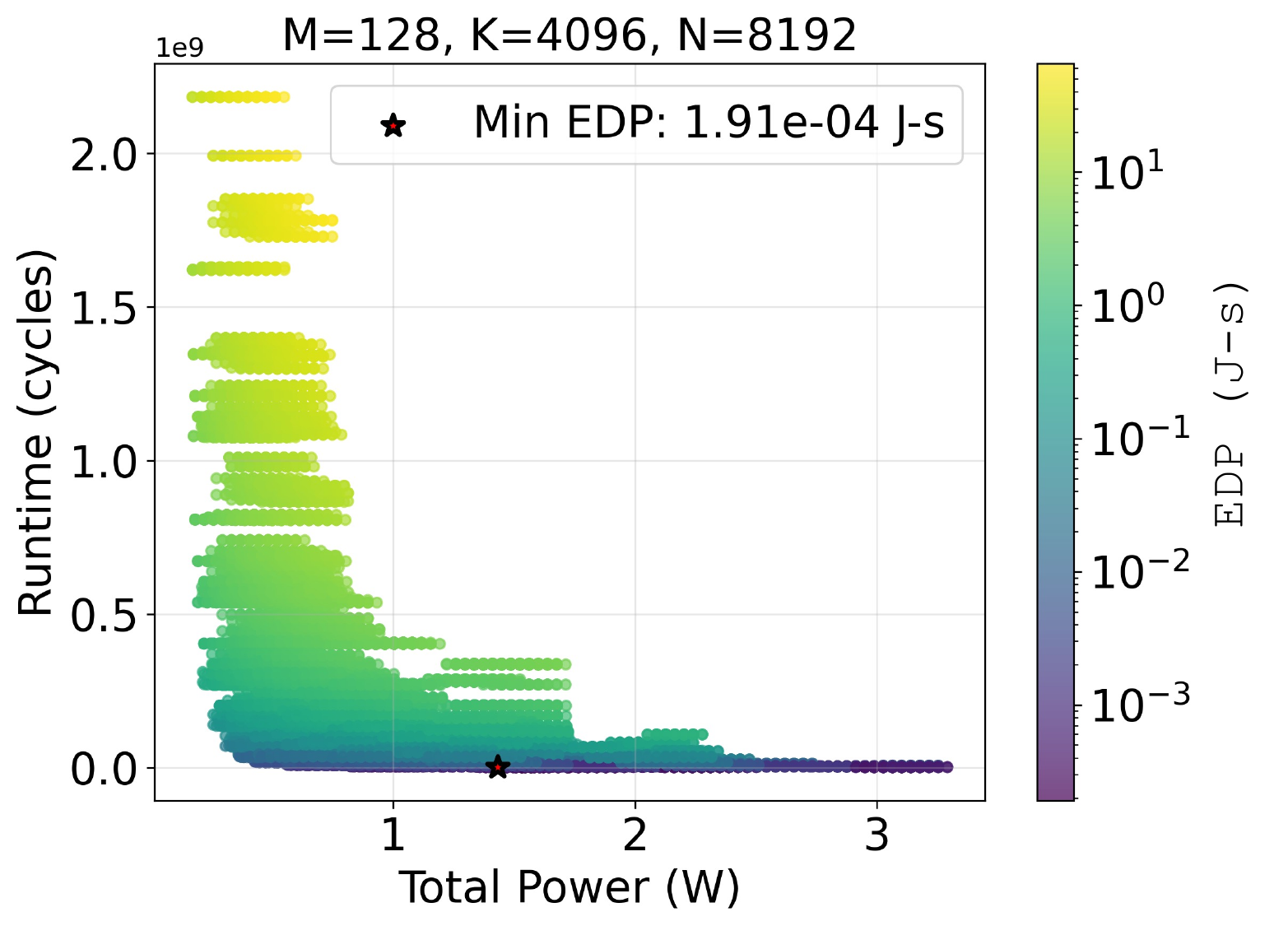}
    \caption{Runtime-power scatter plot for the GEMM operation with dimensions $(M,K,N)$ on a 32~nm ASIC across $7.7\times10^4$ hardware designs sampled from design space of Table~\ref{tab:design_parameters}.}
    \label{fig:edp_trend}
\end{wrapfigure}
 on both power and runtime. The motivation is clear from Fig.~\ref{fig:edp_trend}: diverse hardware designs yield similar EDP values across varied power-runtime tradeoffs, so restricting search to runtime alone may miss configurations with lower EDP. By conditioning DSE on a range of (runtime, power) targets, \ourwork{} can discover low-EDP configurations that would be inaccessible through runtime-only optimization.
To enable this, the Phase-1 PP is extended to jointly predict power and runtime, with loss $\mathcal{L}_{\text{pred}} = (\text{power} - \hat{\text{power}})^2 + (\text{runtime} - \hat{\text{runtime}})^2$ replacing Eqn. 7.
As shown in Fig.~\ref{fig:latent_dse}, the modified $\mathcal{L}_{\text{pred}}$ produces a latent space organized by power-runtime 
class, whereas the raw hardware space lacks any such structure. 
The DDM in Phase 2 is trained using $p= runtime\times power$ as the conditional signal (see Fig.\ref{fig:method}(b)). At inference, for a given EDP target $p$, multiple $HW$ design configurations are generated; we report the lowest EDP $HW$ configuration.

\begin{figure}[h]
    \centering
    \includegraphics[width=0.45\textwidth]{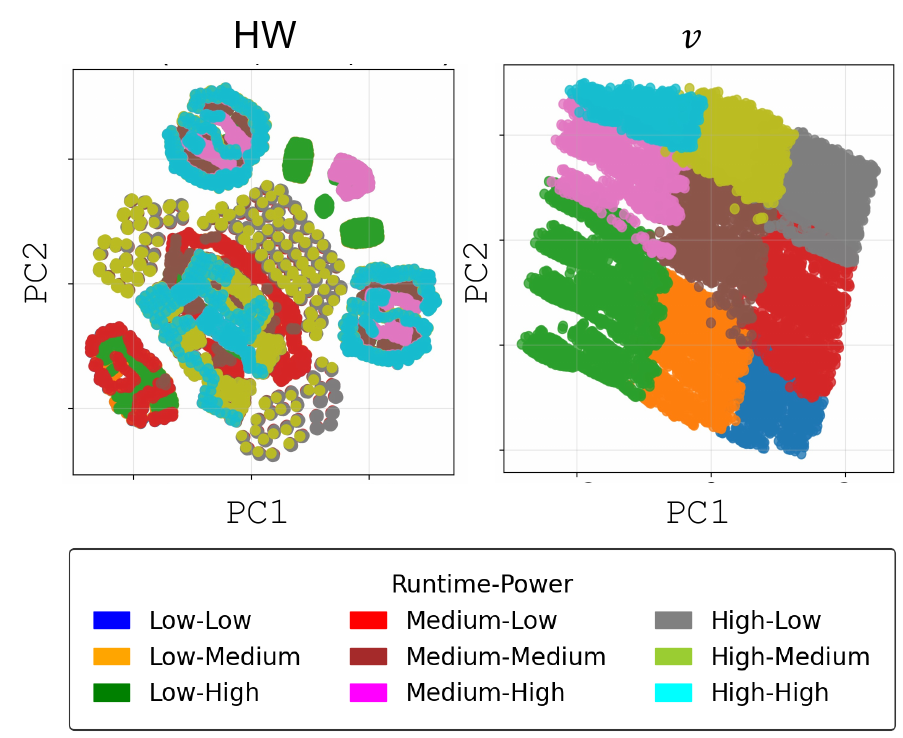}
    \caption{PCA of latent space ($v$) for MLP2 of GPT-2 (decode stage) after Phase-1 training with joint power-runtime loss. 
    The original hardware space (left) lacks semantic structure, while the learned latent space (right) clusters designs by power-runtime 
    class.}
    \label{fig:latent_dse}
\end{figure}

\subsection{DSE for Runtime Optimization}

Fig.~\ref{fig:edp_trend} reveals that lower EDP strongly correlates with lower runtime, suggesting that EDP-conditioned generation can serve as an effective proxy for discovering low-runtime hardware configurations. 
Based on this, for lowest runtime $HW$ configuration generation using DiffAxE, we follow the Phase 1 and Phase 2 training approach as described in Section 4.4 where $p= runtime \times power$ is used for $L_{pred}$ construction and DDM conditioning. 
At inference, generating designs conditioned on (lowest 10\% EDP ranges) yields lowest-runtime configurations that outperform the training data.

\section{Experiments}

\subsection{Dataset}
Each hardware design is represented as a 7-dimensional vector (Table~\ref{tab:design_parameters}). We distinguish two design spaces: the \textbf{training design space}—a coarse-grained subset used to learn the performance-to-hardware mapping—and the \textbf{target design space}—the full fine-grained deployable space. As defined in Table~\ref{tab:design_space_def}, \ourwork{} trains on $O(10^4)$ coarse space and generalizes to $O(10^{17})$ target space. We consider 600 GEMM workloads across varied $M,K,N$ ranges covering varied activation and weight shapes.
\begin{table}[htbp]
\caption{Allowed values of $HW$ Design Parameter Ranges in Training and Target Spaces}
\label{tab:design_space_def}
\centering
\begin{tabular}{@{}p{2cm}p{2.8cm}p{2.8cm}@{}}
\toprule
\textbf{Parameter} & \textbf{Training space} & \textbf{Target space} \\ \midrule
$R$, $C$ & \{4, 8, 16, 32, 64, 128\} & 4--128 (integers) \\
IPSz, WTSz, OPSz & \{4, 64, 128, 256, 512, 1024\} kB & 4--1024 kB (step: 128 bytes) \\
BW & \{2, 4, 8, 16, 32\} B/cycle & 2--32 B/cycle (step: 1 B/cycle) \\
Loop Order & \texttt{mnk}, \texttt{nmk} & \texttt{mnk}, \texttt{nmk} \\
\textbf{Design space} & $7.76\times 10^{4}$ & $5.26\times 10^{17}$\\
\midrule
Workload & \multicolumn{2}{p{0.65\linewidth}}{%
$M: 1$–$1024$, $K: 1$–$4096$, $N: 1$–$30{,}000$ (mean/median: $M \approx 500/500$, $K \approx 800/500$, $N \approx 12000/10000$)
} \\
\bottomrule
\end{tabular}
\end{table} 
Table \ref{tab:design_space_def} outlines the training and generation workloads used in \ourwork{}, covering diverse activation and weight shapes.
The two loop orders for output-stationary (OS) dataflow \cite{du2015shidiannao} are shown for illustration, as OS has been shown optimal across diverse GEMM workloads \cite{samajdar2021airchitect}. Performance labels (runtime, power, EDP) are generated for all $\{HW, w\}$ pairs, yielding $600 \times 7.76\!\times\!10^4 = 46.7$M training points. Runtime is obtained via Scale-Sim \cite{samajdar2018scale}; buffer and DRAM energy via CACTI \cite{balasubramonian2017cacti} at 32~nm; and compute energy via NeuroSim 32nm MAC models \cite{peng2020dnn+} that assume 100MHz clock frequency. All numerical features in $HW$ are min-max normalized. The categorical loop order is one-hot encoded and passed through a learnable linear embedding within the encoder (see Fig. \ref{fig:method} (a), Sec. 4.1). Runtime variability across the design space for a fixed workload is well-documented in accelerator literature \cite{parashar2019timeloop, samajdar2021airchitect}. We also found that for the GEMM workloads considered (Table \ref{tab:design_space_def}), the runtime can span up to three orders of magnitude. To account for this, we apply a logarithmic transformation followed by workload-wise min-max normalization of runtimes to $[0,1]$. Workload dimensions $(M,K,N)$, power and EDP values are similarly min-max normalized.

\subsection{Experimental Setup}

We evaluate \ourwork{} through the following experiments:

\noindent \textbf{Runtime-Specific Hardware Generation}
For each of 600 workloads, 20 target runtimes are uniformly sampled between observed min/max values, yielding 12,000 targets. For each target, 1,000 designs are generated and evaluated via Scale-Sim \cite{samajdar2018scale} ($1.2\times10^7$ total designs). Generation error (lower is better) is defined as $error_{gen} = (T_{gen} - T^*)/T^*$, where $T^*$ is the target runtime. We compare against five baselines: vanilla GD (DOSA~\cite{hong2023dosa}), vanilla BO \cite{reagen2017case}, latent-space GD (Polaris~\cite{sakhuja2024polaris}), latent-space BO (VAESA~\cite{huang2022learning}), and GANDSE~\cite{feng2023gandse}, averaged over 10 random seeds. 
Note, search time (ST) reported in the experiments includes only the optimal $HW$ configuration generation-time search and does not include the one-time training cost for all the methods.

\noindent \textbf{DSE for EDP Optimization}
We introduce \textbf{Search Performance} $SP = EDP_{\text{random}}/EDP_{\text{method}}$ (higher is better) to measure design quality relative to random search. Power and runtime are each binned into three percentile-based categories—low (L: 0–33rd percentile), medium (M: 33–66th percentile), and high (H: 66–100th percentile)—resulting in the joint power-runtime binned into 9 categories: ({HH, HM, HL, MH, MM, ML, LH, LM, LL}).
1,000 designs are generated per category (i.e. 9,000 per workload) and evaluated via Scale-Sim \cite{samajdar2018scale} and CACTI \cite{balasubramonian2017cacti}. Results are averaged over 600 workloads and 10 seeds.

\noindent \textbf{DSE for Runtime Optimization}
We compare against AIRCHITECT~\cite{samajdar2021airchitect}, AIRCHITECT-v2~\cite{seo2025airchitect}, VAESA~\cite{huang2022learning} and GANDSE \cite{feng2023gandse}. For \cite{seo2025airchitect} and \cite{samajdar2021airchitect}, the design space is fixed at 768 as scaling their recommendation-model to $O(10^{17})$ design space is impossible. 
Designs are generated conditioned on lowest 10\% EDP, and the lowest runtime design is selected per workload. All metrics are normalized to \ourwork{}'s generated hardware's runtime and averaged over 600 workloads and 10 seeds.

\section{Results \& Analysis}

\subsection{Runtime-specific Hardware Generation}

\begin{figure}[h]
    \centering
    \includegraphics[width=0.48\textwidth]{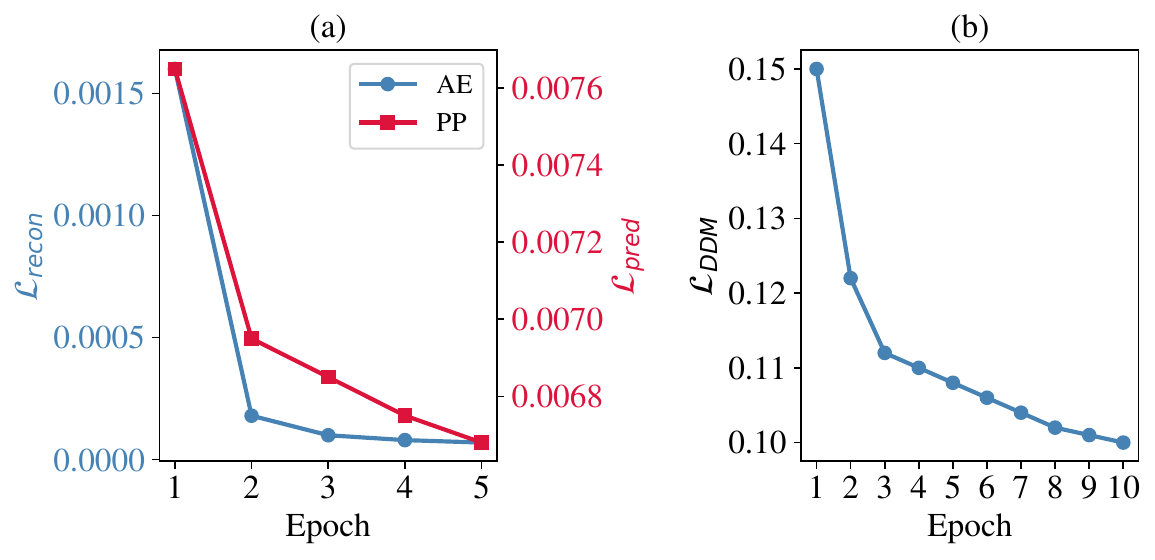}
    \vspace{-20pt}
    \caption{Training loss curves for (a) Phase-1 training of AE+PP: $\mathcal{L}_{pred}$ and $\mathcal{L}_{recon}$ (Eq.6), and (b) Phase-2 training of diffusion model ($\mathcal{L}_{DDM}$ in Eq.2).}
    \vspace{-8pt}
    \label{fig:train_1}
\end{figure}

Fig.~\ref{fig:train_1}(a) shows Phase-1 training curves. The AE and PP are jointly trained for 5 epochs using AdamW \cite{loshchilov2019decoupled} (lr=$10^{-4}$, weight decay=$10^{-3}$) with a \texttt{ReduceLROnPlateau} scheduler (patience=2) and batch size 512 on an NVIDIA V100. Fig.~\ref{fig:train_1}(b) shows Phase-2 DDM training over 10 epochs with AdamW (lr=$10^{-4}$, weight decay=$10^{-2}$), same scheduler, and batch size 128. \ourwork{} comprises only 3.4M trainable parameters and trains in under 7 hours on 46.7M training samples on an NVIDIA V100 GPU.

\begin{table}[h]
\caption{Comparison of \ourwork{} with baselines for runtime-specific hardware generation. $\overline{()}$ denotes average.}
\label{tab:baseline_exp1}
\centering
\resizebox{\columnwidth}{!}{%
\begin{tabular}{@{}lccc@{}}
\toprule
\textbf{Method} & \textbf{Search Time (s)} & \boldmath{$\overline{error}_{gen}$} & \boldmath{$\max\, error_{gen}$} \\ \midrule
Vanilla GD (DOSA~\cite{hong2023dosa}) & 31.54 & 31.59\% & 68.82\% \\
Vanilla BO (BayesOpt) \cite{reagen2017case} & 150.19 & 17.14\% & 59.60\% \\
Latent GD (Polaris~\cite{sakhuja2024polaris}) & 30.80 & 10.08\% & 42.14\% \\
Latent BO (VAESA~\cite{huang2022learning}) & 31.70 & 6.31\% & 41.64\% \\
GANDSE~\cite{feng2023gandse} & $10^{-3}$ & 34.27\% & 70.14\% \\
\textbf{\ourwork{} (ours)} & $1.83\times10^{-3}$ & 5.45\% & 44.86\% \\
\bottomrule
\end{tabular}}
\end{table}

As shown in Table~\ref{tab:baseline_exp1}, vanilla GD yields $\overline{error_{gen}}>30\%$, confirming that differential approximation of $f$ is inaccurate. Vanilla BO reduces error by 13\% over GD but requires $\sim$150\,s. Latent-space methods (GD and BO) outperform their vanilla counterparts, achieving $\overline{error_{gen}}<10\%$ in around 30\,s, validating the utility of the performance-guided latent space. GANDSE achieves ${\sim}32000\times$ speedup over vanilla GD but retains similar error, as both rely on differentiability assumptions. Compared to latent BO \cite{huang2022learning}) with least $\overline{error_{gen}} = 6.31\%$, \ourwork{} achieves lowest average error (5.45\%) at just 1.83\,ms—a $17322\times$ speedup -- demonstrating that diffusion-based latent-space generation outperforms both GD- and BO-based optimization.

\subsection{DSE for EDP Optimization}
As presented in Table \ref{tab:edp_baseline_comparison}, \ourwork{} achieves a search performance (SP) that is 9.8\% higher than VAESA \cite{huang2022learning}, the most competitive baseline employing latent space Bayesian Optimization (BO) within a design space of comparable size and complexity ($O(10^{17})$). 
\begin{table}[h]
\caption{EDP-oriented DSE comparison. SP normalized to random search (↑ better). Search time reported on an NVIDIA V100 GPU.}
\centering
\resizebox{\columnwidth}{!}{%
\begin{tabular}{@{}lccc@{}}
\toprule
\textbf{Method} & \textbf{Design Space} & \textbf{SP~(↑)} & \textbf{Search Time~(↓)} \\
\midrule
Random Search & $O(10^{17})$ & 1.00 & 3 ms \\
Vanilla BO(BayesOpt)~\cite{reagen2017case} & $O(10^{17})$ & 0.98 & 2 hours \\
VAESA~\cite{huang2022learning} & $O(10^{17})$ & 1.02 & 40 min \\
DOSA~\cite{hong2023dosa} & ${\sim}O(10^{7})$ & 0.20 & 18 min \\
Polaris~\cite{sakhuja2024polaris} & ${\sim}O(10^{7})$ & 0.54 & 10 min \\
\textbf{\ourwork{} (ours)} & $O(10^{17})$ & \textbf{1.12} & \textbf{16.5 s} \\
\bottomrule
\end{tabular}}
\label{tab:edp_baseline_comparison}
\end{table}
Importantly, \ourwork{} attains this superior SP within a fixed search time of only 16.5 seconds across 9 categories of H,M,L power-runtime ranges, resulting in a \textbf{145.6$\times$} improvement in search speed. In contrast, DOSA \cite{hong2023dosa} and Polaris \cite{sakhuja2024polaris}, which rely on vanilla and latent space gradient descent (GD) respectively, yield significantly suboptimal designs—exhibiting lower SP (i.e., higher EDP) than even random search. This underperformance is primarily attributed to their limitation to searching in a coarser and smaller design space (approximately $O(10^{7})$), limiting their optimization effectiveness.

\subsection{DSE for Runtime Optimization}
\begin{figure}[h]
    \centering
    \includegraphics[width=0.5\textwidth]{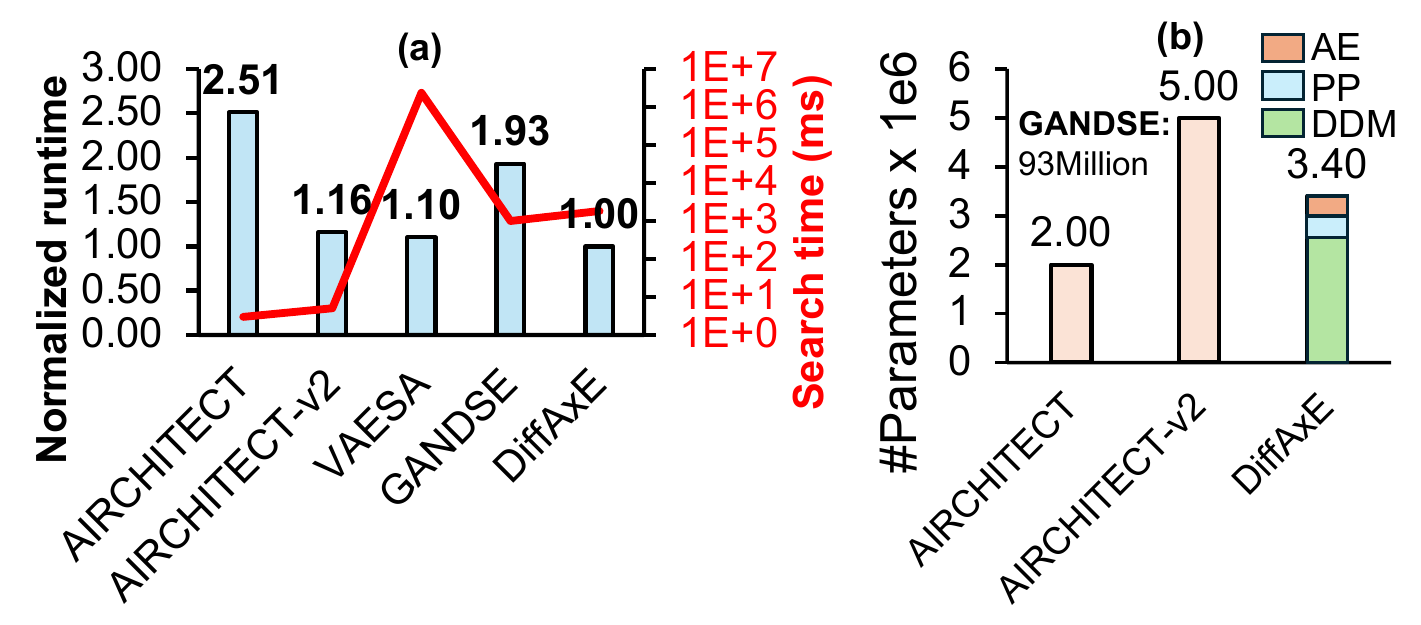}
    \vspace{-16pt}
    \caption{(a)Normalized runtime and GPU search time of baselines compared to \ourwork{}. (b)Comparison of model size between prior DL-based DSE works and \ourwork{} with component-wise breakdown. All metrics are lower the better.}
    \label{fig:perf_speedup}
\end{figure}
Fig.~\ref{fig:perf_speedup}(a) presents the runtime of designs generated by \ourwork{} compared to prior works. When compared to VAESA, which employs latent-space BO, \ourwork{} achieves a 10\% lower runtime while reducing search time by a factor of \textbf{1312$\times$}. Relative to the DL-based classification approach used in AIRCHITECT, \ourwork{} delivers an average runtime improvement of \textbf{2.51$\times$}. Furthermore, in comparison to AIRCHITECT v2 \cite{seo2025airchitect}, \ourwork{} achieves a 16\% lower runtime. Further, \ourwork{} requires 32\% lower number of parameters compared to AIRCHITECT v2 as shown in Fig.~\ref{fig:perf_speedup}(b). Notably, \ourwork{} also achieves \textbf{1.93$\times$} lower runtime configurations than GANDSE \cite{feng2023gandse} at \textbf{27$\times$} lower parameter count, demonstrating that performance-aware latent clustering enables a significantly more compact and accurate generative model than the GAN-based surrogate approach.

\section{Inference Optimization of LLMs}
\ourwork{} extends to LLM inference EDP minimization across prefill and decode phases, which have distinct computational profiles motivating specialized per-phase accelerator configurations in chiplet-based systems \cite{krishna2018chiplet, heo2023chiplet, jain2023multichip, wang2025hisim}. To generalize from single GEMM workloads to full DNNs comprising sequences of GEMMs, the MLP-based PP (Fig. 3(a)) is replaced with an attention-based sequence encoder that captures inter-layer dependencies.
\begin{figure}[h]
    \centering
    \vspace{-4pt}
    \includegraphics[width=0.5\textwidth]{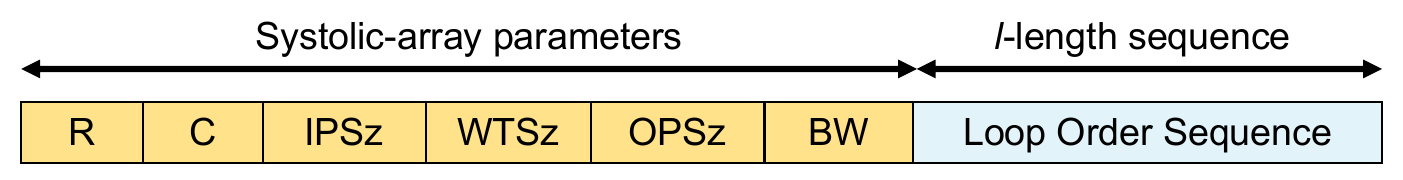} 
    \caption{Data structure of $HW$ in Fig. \ref{fig:method}(a) for sequence of $l$ GEMM operations across l-layers of LLM. The loop order sequence represents the independent mapping of each layer.}
    \label{fig:data_struc_seq}
    \vspace{-6pt}
\end{figure}
\begin{figure*}[h]
    \centering
    \includegraphics[width=\textwidth]{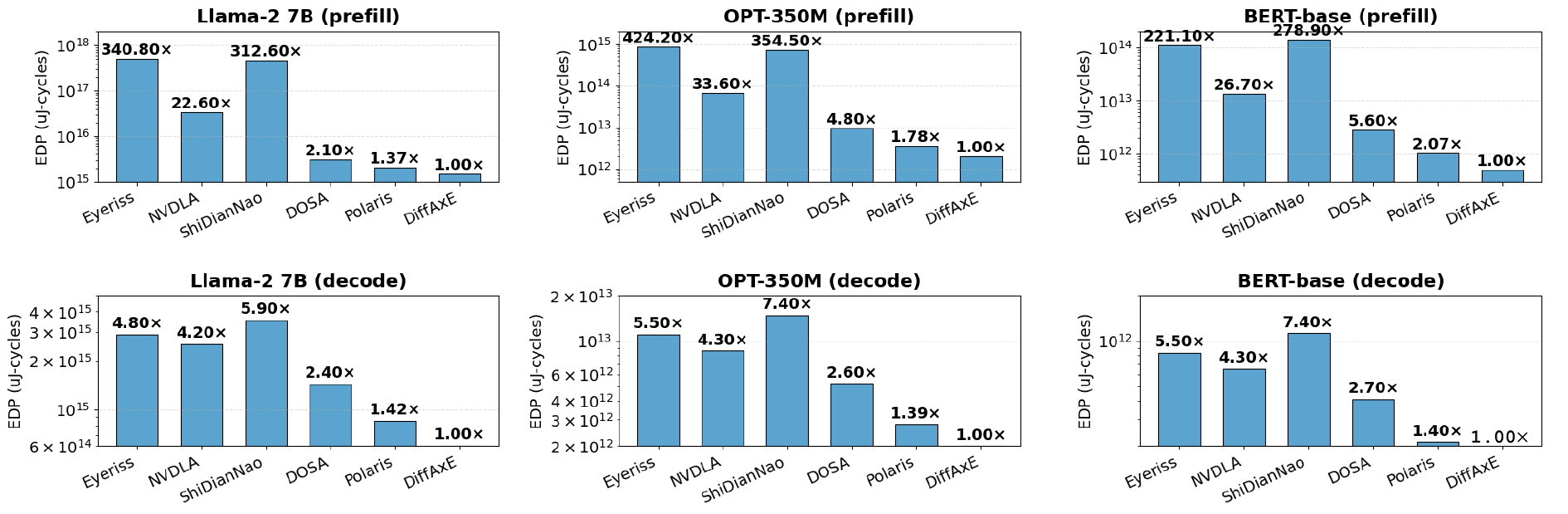} 
    \vspace{-20pt}
    \caption{Energy-delay product (EDP) of baseline accelerators including Eyeriss \cite{chen2016eyeriss}, NVDLA \cite{zhou2018research}, ShiDianNao \cite{du2015shidiannao}, DOSA \cite{hong2023dosa}-optimized Gemmini \cite{genc2021gemmini} and Polaris \cite{sakhuja2024polaris} compared to \ourwork{} for 32nm ASIC implementation. Prefill represents a default sequence length of 128 tokens. Bar labels represent EDP normalized to \ourwork{}.}
    \label{fig:llm_edp}
    \vspace{-8pt}
\end{figure*}
The $HW$ representation supports layer-wise loop order generation while keeping systolic array parameters fixed across layers (Fig.~\ref{fig:data_struc_seq}). 
\begin{wrapfigure}{r}{0.20\textwidth}
    \vspace{-1em}
    \centering
    \includegraphics[width=0.20\textwidth]{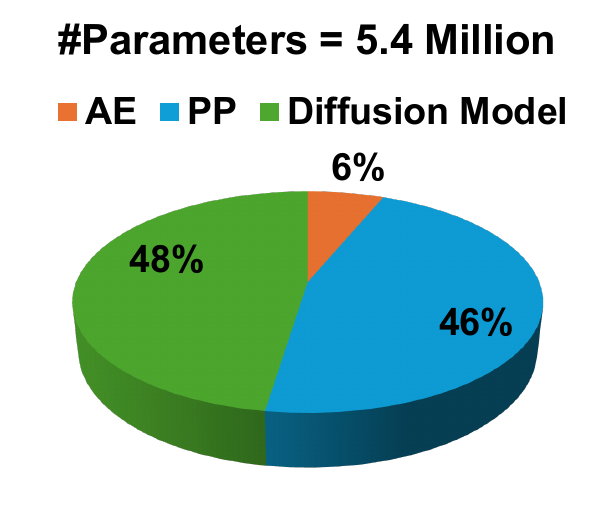}
    \vspace{-20pt}
    \caption{Model size of \ourwork{} for GEMM sequence workloads and component-wise breakdown.}
    \label{fig:params_seq}
    \vspace{-1em}
\end{wrapfigure}
This extension increases model size to 5.4M parameters due to the attention-based PP (Fig. 9), but our evaluations indicate generation time stays largely unaffected as the DDM architecture and denoising steps remain unchanged. We evaluate \ourwork{} on LLaMA-2 7B \cite{touvron2023llama}, OPT-350M \cite{zhang2022opt}, and BERT-base \cite{devlin2019bert}, comparing against fixed accelerators -- Eyeriss \cite{chen2016eyeriss}, NVDLA \cite{zhou2018research}, ShiDianNao \cite{du2015shidiannao}, and DSE works -- DOSA \cite{hong2023dosa} and Polaris \cite{sakhuja2024polaris}. 
Fig.~\ref{fig:llm_edp} shows that \ourwork{} and DOSA consistently outperform fixed architectures across all LLMs and stages, with the largest gains during prefill due to greater flexibility in PE array sizing and DRAM bandwidth. Notably, \ourwork{} achieves more than $2\times$ lower EDP than DOSA and over $1.4\times$ lower EDP than Polaris across all LLMs and stages.
\begin{table*}[h]
\centering
\caption{Hardware design comparison: DOSA vs.\ \ourwork{} on BERT workloads. Loop order shown per layer in [].}
\label{tab:deep_dive_dosa}
\resizebox{\textwidth}{!}{%
\begin{tabular}{llccccccccc}
\toprule
\textbf{Workload} & \textbf{Method} & $R$ & $C$ & IPSz & WTSz & OPSz & BW & \textbf{Loop Order} & \textbf{Runtime} & \textbf{EDP} \\
& & & & (kB) & (kB) & (kB) & (B/cycle) & [qkv, $qk^{T}$, sv, attn\_proj, mlp1, mlp2] & (cycles) & (uJ-cycles) \\
\midrule
\multirow{2}{*}{BERT-Base-prefill}
& DOSA & 128 & 128 & 128 & 128 & 64 & 32 & [mnk, mnk, mnk, mnk, mnk, mnk] & 1.42e+07 & 2.80e+12 \\
& \ourwork{} & 128 & 63 & 1024 & 4 & 8.5 & 32 & [nmk, mnk, mnk, nmk, nmk, nmk] & 6.90e+06 & \textbf{5.02e+11} \\
\midrule
\multirow{2}{*}{BERT-Base-decode}
& DOSA & 128 & 128 & 96 & 128 & 32 & 32 & [nmk, mnk, mnk, nmk, nmk, mnk] & 5.82e+06 & 4.71e+11 \\
& \ourwork{} & 4 & 64 & 4 & 4 & 8.9 & 32 & [mnk, mnk, mnk, mnk, mnk, nmk] & 5.47e+06 & \textbf{1.75e+11} \\
\bottomrule
\end{tabular}}
\end{table*}
Table~\ref{tab:deep_dive_dosa} provides a detailed hardware configuration comparison with DOSA for BERT-base. During prefill, \ourwork{} selects \texttt{nmk} ordering and maximizes input buffer (1~MB) to minimize redundant DRAM accesses, whereas DOSA's smaller buffer allocation incurs frequent redundant fetches — resulting in $>2\times$ longer runtime and $\sim6\times$ higher EDP. During decode (sequence length of 1 token), where $M=1$ across all layers, \ourwork{} selects small $R$ to avoid PE under-utilization and output drain overhead, reducing buffer requirements and dynamic power. Despite comparable runtimes, DOSA's larger buffers and PE under-utilization lead to $\sim3\times$ higher EDP.

\noindent \textbf{FPGA Implementation:} Next, we implement the accelerator designs for BERT-base prefill stage on a Xilinx Virtex UltraScale+ VU13P FPGA \cite{xilinx_ds890_2022, xilinx_virtexUSplus_brief_2025, xilinx_power_efficiency_2025}. 
Table \ref{tab:fpga_res} shows the component-wise resource utilization for different accelerator designs on the FPGA. 
\begin{table}[htbp]
\centering
\caption{Resource utilization of various DNN accelerator architectures. DOSA and DiffAxE shown for Bert-Base-prefill case (Table \ref{tab:deep_dive_dosa}).}
\label{tab:fpga_res}
\resizebox{\columnwidth}{!}{%
\begin{tabular}{lccccc}
\toprule
\textbf{Architectures} & \textbf{\#DSP} & \textbf{\#LUT} & \textbf{\#FF} & \textbf{\#BRAM} & \textbf{\#URAM} \\
\midrule
Eyeriss~\cite{chen2016eyeriss} & 84 & 45,696 & 71,544 & 10 & 6 \\
ShiDianNao~\cite{du2015shidiannao} & 128 & 47,632 & 74,448 & 26 & 0 \\
NVDLA~\cite{zhou2018research} & 512 & 64,528 & 99,792 & 31 & 15 \\
DOSA~\cite{hong2023dosa} & 8192 & 360,448 & 540,672 & 23 & 8 \\
\textbf{DiffAxE} & 4032 & 232,408 & 352,112 & 11 & 29 \\
\bottomrule
\end{tabular}%
}
\end{table}
For DOSA \cite{hong2023dosa} and \ourwork{}, we report utilization for the architectures generated for BERT-Base prefill stage (see Table \ref{tab:deep_dive_dosa}). As expected, DOSA and \ourwork{} designs require much larger DSPs due to higher number of MAC units compared to the fixed architectures like NVDLA. Also, \ourwork{} consumes highest amount of URAM (UltraRAM) due to largest aggregate on chip buffer size as shown in Table \ref{tab:deep_dive_dosa}. 
\begin{figure}[h]
    \centering
    \includegraphics[width=0.48\textwidth]{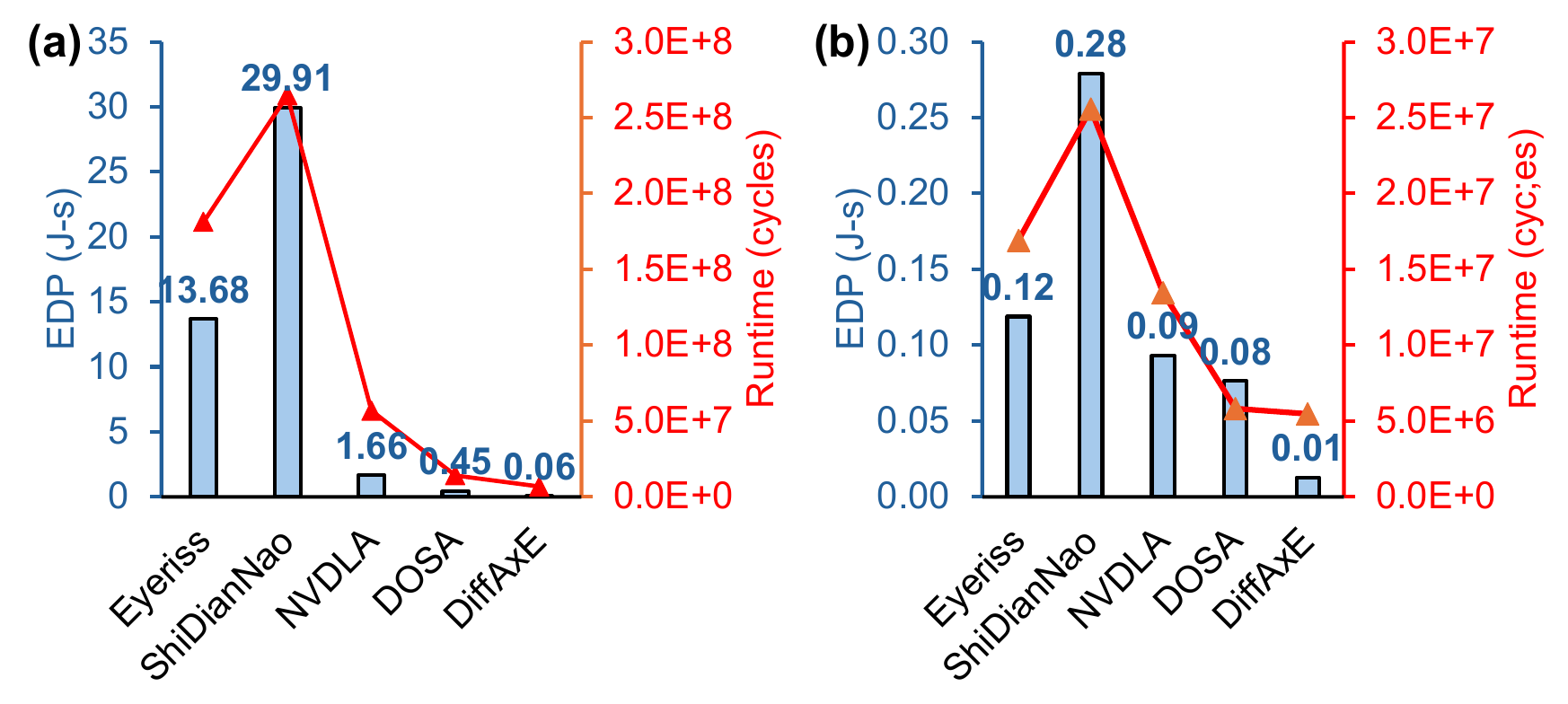} 
    \vspace{-16pt}
    \caption{EDP and runtime of different architectures for BERT-base (a) prefill(128 token length) and (b) decode (1 token generation) stages on Xilinx Ultrascale+ VU13P FPGA.}
    \label{fig:fpga_power}
    \vspace{-12pt}
\end{figure}
Fig. \ref{fig:fpga_power} shows the EDP of BERT-base prefill and decode stages during inference using the three fixed architectures, and two optimization based strategies--DOSA and \ourwork{}. Clearly, \ourwork{} achieves the lowest EDP among all the designs, with a \textbf{7.5x} and \textbf{8x} improvement over DOSA during prefill and decode stages respectively.

\section{Conclusion}
We introduce \ourwork{}, a diffusion based generative framework for performance-conditioned accelerator design and design space exploration (DSE) scaling to exponentially large design spaces.
Overall, \ourwork{} enables efficient hardware generation and optimization with significantly reduced search cost, while achieving iso- or better performance than prior methods. While \ourwork{} currently targets single-core accelerators, future work can extend this framework to tiled multi-core architectures by adopting tree- and graph-based hardware encoding \cite{gong2025crane, cai2023inter}.


\bibliographystyle{ACM-Reference-Format}
\bibliography{sample-base}
%

\end{document}